\newcommand{\nop}[1]{}
\documentclass[conference]{IEEEtran}
\makeatletter
\def\ps@headings{%
\def\@oddhead{\mbox{}\scriptsize\rightmark \hfil \thepage}%
\def\@evenhead{\scriptsize\thepage \hfil \leftmark\mbox{}}%
\def\@oddfoot{}%
\def\@evenfoot{}}
\makeatother
\ifCLASSINFOpdf
  \usepackage[pdftex]{graphicx}
  \DeclareGraphicsExtensions{.jpeg,.png}
\else
  \usepackage[dvips]{graphicx}
   \DeclareGraphicsExtensions{.eps,.ps}
\fi
\usepackage[caption=false,font=footnotesize]{subfig}

\usepackage{amsthm}
\usepackage{amssymb}

\providecommand{\expectation}[2]{\mathbb{E}_{#2}\left[#1\right]}
\providecommand{\probab}[2]{\mathbb{P}_{#2}\left\{#1\right\}}

\providecommand{\vcprob}[3]{f^{#1}_{#2}(#3)}
\providecommand{\rktree}[2]{G^{#1}(#2)}
\providecommand{\ktree}{$k$-tree}
\providecommand{\ktrees}{$k$-trees}

\providecommand{\binom}[2]{{#1\choose#2}}

\providecommand{\degree}[2]{{\textrm{deg}_{#1}(#2)}}

\renewcommand{\qed}{\hfill $\blacksquare$}

\renewcommand{\theequation}{\thesection.\arabic{equation}}

\newtheorem{definition}{Definition}
\newtheorem{theorem}{Theorem}
\newtheorem{lemma}{Lemma}

\begin{document}
\title{Statistical Behavior of Embeddedness and Communities of Overlapping Cliques in Online Social Networks}
\author{\IEEEauthorblockN{Ajay Sridharan}
\IEEEauthorblockA{University of \\
Victoria,\\
Email: ajays@uvic.ca}
\and
\IEEEauthorblockN{Yong Gao}
\IEEEauthorblockA{University of \\
British Columbia,\\
Email: yong.gao@ubc.ca}
\and
\IEEEauthorblockN{Kui Wu}
\IEEEauthorblockA{University of \\
Victoria,\\
Email: wkui@ieee.org}
\and
\IEEEauthorblockN{James Nastos}
\IEEEauthorblockA{University of \\
British Columbia,\\
Email: jnastos@interchange.ubc.ca}
}


%Telephone: (800) 555--1212, Fax: (888) 555--1212}


\maketitle

\begin{abstract}
Degree distribution of nodes, especially a power law degree distribution, has been regarded as one of the most significant
structural characteristics of social and information networks. Node degree, however, only discloses the first-order structure
of a network. Higher-order structures such as the edge embeddedness and the size of communities  may play
more important roles in many online social networks.
\nop{These higher-order structures have not been systematically analyzed before.}
In this paper, we provide empirical evidence on the existence of rich higher-order structural characteristics in online social networks,
develop mathematical models to interpret and model these characteristics, and discuss their various applications in practice. In particular,
\begin{enumerate}
\item We show that the embeddedness distribution of social links in many social networks has interesting and rich behavior that
cannot be captured by well-known network models. We also provide empirical results showing a clear correlation between the embeddedness distribution and the average number of messages communicated between pairs of social network nodes.
\item We formally prove that random $k$-tree, a recent model for complex networks, has a power law embeddedness distribution, and
show empirically that the random $k$-tree model can be used to capture the rich behavior of higher-order structures we observed
in real-world social networks. \nop{To the best of our knowledge, this is the first existing model for which a power law distribution has
been established mathematically for non-trivial structural measures of a network other than the node degree.}
\item Going beyond the embeddedness, we show that a variant of the random $k$-tree model can be used to
capture the power law distribution of the size of communities of overlapping cliques discovered recently.
\nop{by Palla et al.~\cite{Palla}}
\end{enumerate}

\end{abstract}

\IEEEpeerreviewmaketitle

\section{Introduction} \label{sec:introduction}

Social networks essentially represent the relationship between two social entities such as individuals or organizations.
It indicates the way in which they are connected, ranging from casual acquaintance to personal connections.
The advent of the Internet has revolutionized the way of communication amongst the common masses.
One of the popular services that spawned out of this revolution is the On-line Social Network (OSN),
evidenced by the huge success and popularity of On-line Social Network (OSN) websites such as Facebook,
MySpace and Twitter, all having hundreds of millions of users. These OSNs not only provide
their users with a convenient environment to interact easily with their friends, colleagues, relatives,
and even ``strangers" who share common interests, but also serve as a mirror of real social networks,
making the study of social structure and interaction much easier than before. As a result, the study on
the structural behavior of on-line social networks has triggered unprecedented interests in many research areas,
including telecommunication networks, social science, and business to mention a few~\cite{Easley10}.
It is also one of the core research problems in the newly emerging scientific discipline, network science~\cite{Lewis09}.

The power law distribution of node degree has been regarded as one of the most significant structural characteristics of
social and information networks. In 1999, Barab\'{a}si and Albert discovered that the degree distribution of the
World Wide Web (WWW) follows a power law~\cite{barabasi-1999-286}. Since then, this structural behavior
has been broadly investigated in many other types of real-world networks~\cite{Easley10} and a large variety of generative
random models have been proposed for it~\cite{Durrett07}. Nevertheless, node degree distribution alone only discloses the first-order structure of such networks. Higher-order structures such as the edge embeddedness, a notion used to capture the
``degree" of a social tie with regard to the number of common neighbors, and the size of communities
may play a more important role in information propagation and on-line social networking.

\nop{
%In many cases, we should care more about how close the tie is between two online entities (i.e., the degree of a social tie),
%and the popularity of a node is of the second-order importance. To make the argument intuitive, consider Bill Gates on
%Facebook and Twitter. It is not surprising that he has so many ``friends" in both OSNs~\cite{Gaudin10}.
%In this sense, Bill Gates becomes a very popular node in the OSNs and the node degree of this node is extremely high.
%Despite the node's popularity, however, it would be unwise to send Bill Gates a message about your university
%admission to MIT, certainly a great news to you, and hope the message being propagated to a large group of audience.
}

Practically, the degree of a social tie has already been utilized in various application contexts. As an incomplete list,
Zhu et al. use the ``cellular-social relationship graph" constructed based on traffic amount between cellular users in
the design of effective patching strategy to prevent the propagation of computer worms over smart phones~\cite{Zhu09};
Ioannidis and Massoulie use the implicit tie between friends, i.e., the shared common interests, to develop personalized
strategies for searching the web~\cite{Ioannidis10}; Wolf et al. use social network analysis on the communication structure
of development teams, another type of social tie between network nodes, to predict software build failures~\cite{Wolf09}.

Despite the diverse practical applications, existing work is mostly based on empirical studies over real-world dataset~\cite{Wolf09,Palla05,mislove-2007-orkut,viswanath-2009-fb}. While empirical studies are valuable, their pitfalls are obvious: the data collection is time consuming; the size of dataset is usually huge and very hard to handle; the cost of human resources on analyzing and processing the data is nontrivial. As mathematical models can greatly alleviate the above problems, the call for new mathematical models that are powerful and flexible enough to capture the statistical behavior of OSNs, especially the higher-order statistics like the embeddedness, becomes unprecedentedly urgent.

While there are numerous mathematical models designed to model the structural behavior of complex networks~\cite{Easley10,barabasi-1999-286,jonasson99jap,snijders06sociamethod}, to the best of our knowledge, there is currently no unified mathematical framework to design \textit{generative models that are able to model the statistical behavior of higher-order structures such as the embeddedness or communities.} In this paper, we address the above challenge with the following contributions:
\begin{enumerate}
\item We show that in some real-world OSNs like Facebook, the distribution of edge embeddedness,
a notion used to capture the ``degree" of a social tie with regard to the number of common neighbors,
has interesting and rich behavior that cannot be captured by well-known network models designed
to model the observed power law node degree distribution in information networks.
We also provide empirical evidence showing a clear correlation between a power law embeddedness distribution and
the average number of messages communicated  between pairs of social network nodes.
\item We prove formally that random $k$-tree, a recent model for complex networks~\cite{Gao09}, has a power law distribution of
embeddedness. We show empirically that random $k$-trees can be used to model and interpret the statistical behavior of embeddedness
we have observed in real-world social networks.
To the best of our knowledge, this is the first existing model for which a power law distribution has
been established mathematically for higher-order structural measures of a network other than the node degree.
\nop{Together with the existing proof that random $k$-tree creates a network having power law distribution of node degree,
our new results provide strong evidence that random $k$-tree model can serve as a very good approximate model for real-world OSNs.}
\item Going beyond the embeddedness, we show that a variant of
the random $k$-tree model generates random networks that
capture well the power law distribution of the size of communities of
overlapping cliques as has been discovered by Palla et al.~\cite{Palla05} in 2005.
\nop{\item We point out the significance of embeddedness to motivate further applications using OSNs.}
\end{enumerate}

\section{Background} \label{sec:background}
\subsection{Structural Properties of OSNs}

An OSN is usually modeled as an undirected graph $G= G(V, E)$ where $V$ denotes the set of nodes and $E$ denotes the set of edges.
A node represents an individual entity (e.g., a people or an organization) and an edge between two nodes signifies a
social connection between the individual entities established according to some given criteria such as friendship or colleagues.
In graph theory, a node is also called a vertex, and in the sequel we will use the two terms interchangeably.
\nop{An edge between two vertices $u$ and $v$ is denoted by $e=\{u,v\}$.}Let $e=\{u,v\}$ denote an edge between the two nodes $u$ and $v$. The degree of a vertex $v$ in a network $G$, denoted by
$\degree{G}{v}$, is the count of its neighbors. Among the many structural properties of OSNs~\cite{Easley10}, the distribution of vertex degrees of a network is probably the most well-known one that has been broadly studied before.
\begin{definition}
\textbf{Power law distributions}. A power law distribution, as the name suggests,  is a distribution function of the form
$
  F(x) = 1 - x^{-\alpha + 1} \textrm{ for some constant }  \alpha > 1.
$
The corresponding density function is $f(x) = -c x^{-\alpha}$ for $c = -\alpha + 1$. Note that $\alpha$ is also called the power law exponent. 
\end{definition}

Power law distributions have long been used as a tool to model and explain empirical observations in a large variety of
research fields. A distinct feature of a power law distribution is that it has a ``heavy tail" as compared to other well-known distributions such as
the normal distribution, the Poisson distribution and the exponential distribution. People
are interested in such a distribution because it is scale free, i.e., scaling the variable $x$ does not change the shape
of the function. \nop{i.e., letting $y = \frac{x}{t}$, we have
$f(y) = \frac{c\alpha}{t^{-\alpha}}x^{-\alpha} = c_1 y^{-\alpha}$ where $c_1 = c\alpha$.}

The degree sequence of a network $G$ is a sequence of integers $\{d_1, \cdots, d_n\}$ where $d_i$ is the degree of the $i$-th vertex.
The degree distribution of a network $G$ is a sequence $\{X_1, \cdots, X_n\}$ where $X_d$ is the proportion of
vertices with degree $d$. It has been discovered that the degree distribution of the World Wide Web (WWW) and many other real-world networks follows
a power law distribution~\cite{barabasi-1999-286,Easley10}. To interpret their empirical observations,
Barab\'{a}si and Albert proposed an evolving random model  that is known as
the \textit{preferential attachment model} (also called \textbf{the BA model}).\nop{Simply put, this model defines a
graph evolution process in which vertices are added to the graph one at a time.} According to this model, a graph grows by adding one vertex at a time. In each step, the newly-added vertex is connected to $m$ existing vertices selected according to the preferential attachment mechanism, i.e.,
an existing vertex is selected with probability in proportion to its degree.

Bollob\'{a}s et al. \cite{bollobas01scalefree} later provided a formal proof showing that the vertex degree distribution of
the BA model obeys a power law distribution~\cite{bollobas01scalefree}. \nop{More formally, it is proved in \cite{bollobas01scalefree} that
with high probability the proportion of vertices with a given vertex degree $d$
is asymptotically equivalent to $d^{-3}$.}Since then, quite a few similar models have been proposed and studied
\nop{(\cite{aiello01,kumar00,cooper03,jordan06})}
\cite{cooper03} in order to  design models where the scaling exponent of the power law
vertex degree distribution can be controlled by some parameters. However, we note that \textit{none} of these models are designed to capture structural characteristics other than the vertex degree distribution and clustering coefficients.

As the vertex degree distribution only  provides statistics of the degree of individual vertices in a network, we regard
it as the first-order structural property. In graph theory, it is well known that the structure of a network can
be fully characterized by its vertex degree distribution only if the network belongs to some very special class of graphs.
To get a more general picture about the structural features
of an OSN, we may in many cases care more about the statistics of other higher-order structural
properties such as the embeddedness and communities, which are the main focus of this paper.

\begin{definition}
\textbf{(Edge) Embeddedness}. The \textbf{embeddedness} of an edge $e = \{u, v\}$ in a network $G(V, E)$, denoted by $\degree{G}{e}$,
is defined to be the number of common neighbors of $u$ and $v$. For an edge $e = \{u, v\}$ with $\degree{G}{e} = d$,
the subnetwork consisting of the vertices $u$, $v$, and their $d$ common neighbors is called a $d$-triangle.
%It is also called the \textbf{edge embeddedness} of degree $d$.
\end{definition}

\begin{figure}[!t]
\centering
\includegraphics[width=1in]{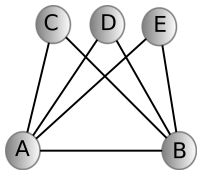}
\caption{Edge embeddedness of degree 3}
\label{fig_dt}
\end{figure}

%Note that edge embeddedness of degree $d$ in some cases is called $d$-triangles as well, because associated with the edge are
%$d$ triangles.
Embeddedness, also called edge embeddedness interchangeably, is an important measure of social networks~\cite{snijders06sociamethod}. Fig.~\ref{fig_dt} shows an example of a $3$-triangle where
the edge $AB$ has embeddedness of degree $3$. Edge embeddedness can be regarded as the ``degree" of an edge. This is the reason why we use $\degree{G}{e}$ to denote
the embeddedness of an edge.

In addition to edge embeddedness, many OSNs typically include \nop{subsets}subgroups of nodes that are \nop{highly}strongly
connected to each other than to the rest of the network. Such \nop{subsets}subgroups are generally called communities, even though no unique
definition is widely acceptable so far. As demonstrated in~\cite{Palla05}, statistics of communities and their overlapping disclose
interesting social/natural relationships in many real networks. We follow the definition in~\cite{Palla05}.

\begin{definition}
\textbf{Community of Overlapping Cliques}. A $k$-clique community in a network is defined as a union of all $k$-cliques
(i.e., complete subgraph of size $k$) that can be reached from each other through a series of adjacent $k$-cliques,
where adjacency means sharing $k-1$ nodes.
\end{definition}

\nop{In the rest of the paper, we will mainly focus on the above three structure properties of OSNs: distribution
of node degree, distribution of embeddedness, and distribution of the size of communities of overlapping $k$-cliques.}

\nop{With the discovery of power law distribution in Internet and WWW by Barabsi et al. [citation] many graph models
have been proposed incorporating power law distribution as an important factor in their design.
The study of structure of networks has a long history \cite{Bollobas} in the field of mathematics
which also includes the seminal work in Random graph models by outstanding mathematicians Erdos and Renyi\cite{ErdRn60}.
The concept of small-world in many networks where the distance between any two nodes is relatively
short bears its roots to the work done by the social psychologist Stanley Milgram\cite{milgram}.
The graph generation model introduced by Watts and Strogatz [j1998collective] has the properties of a
small world network which include high clustering coefficient and shortest-path between nodes.
In \cite{{albert99a}, {huberman}, {Kleinberg99theweb}} the network of the World Wide Web and its links were
studied which led to interesting results that showed an evidence of a power-law distribution.
It also showed that most real world networks are scale-free in nature.
While there are numerous network models in the literature, most of them can be classified under
three main categories which include simple/random network, highly clustered or connected network and scale-free network.}

\subsection{Random $k$-Tree Model}

In spite of the diverse applications of higher-order structural properties like embeddedness and community in OSNs,
there are currently no generative mathematical models amenable for deriving the distribution of these higher-order structural
properties. In~\cite{Gao09}, one of the authors of this paper introduced a random $k$-tree model and proved that
the random $k$-tree model generates graphs with vertex degree distribution following a power law. To ease our further discussion,
we introduce this model first.

%Comments: The brief introduction comes here.
\nop{Recall that the degree of a vertex in a graph $G$ is denoted by $\degree{G}{v}$.}
\nop{Throughout this paper, a $k$-clique of a graph is understood as a complete subgraph on $k$ vertices.}Throughout this paper, a fully connected subgraph of $k$-vertices is defined to be a $k$-clique.
All the graphs considered  are undirected.
The construction of a random \ktree\ is based on the following simple randomization of the recursive definition of \ktrees .
Starting with an initial $k$-clique $\rktree{k}{n}$, a sequence of graphs $\{\rktree{k}{n}, n \geq k\}$ is constructed
by adding vertices to the graph, one at a time. To construct $\rktree{k}{n + 1}$ from $\rktree{k}{n}$, we add a new vertex
$v_{n + 1}$ and connect it to
the $k$ vertices of a $k$-clique selected uniformly at random from all the $k$-cliques in $\rktree{k}{n}$.
We call the graph process $\{\rktree{k}{n}, n \geq k\}$ a \textit{$k$-tree process}.

\nop{Intuitively, the random $k$-tree model may behave not exactly the same way as people join OSNs. For example, after Alice receives
an invitation from Bob to join an OSN, Alice may only establish connections to part of Bob's friends. Nevertheless,
random $k$-tree model can approximate the network creation process with a properly-chosen value of $k$. Most importantly, this model
has nice mathematical structures making theoretical analysis of higher-order structure properties of OSNs easier.}

\section{The Edge Embeddedness Distribution and Contact Strength of Social Links in OSNs}
%\Ajay's section/changes starts here, Kui has revised it.

%In this section, we present the empirical study on distribution of node degree and embeddedness in real-world OSNs.
%We also demonstrate that the random $k$-tree model and one of its variants can capture the statistical behavior of real-world OSNs.

While the vertex degree distribution of real-world networks has been intensively studied since the seminal work of Barab\'{a}si and Albert
\cite{barabasi-1999-286}, we know of no previous work on the statistical behavior of the edge embeddedness. Our work has been motivated by the belief that an
understanding on the statistical behavior of higher-order structures, such as the embeddedness, may help shed further light on the structure
and the dynamics of OSNs. In this section, we report our empirical studies focusing on
\begin{enumerate}
\item the distribution of edge embeddedness in OSNS;
\item the impact of embeddedness distribution on social communication;
%\item the possible mathematical models that can capture OSN's statistical behavior beyond the degree distribution.
\end{enumerate}
We will discuss the possible models that can capture the observed behavior of  the edge embeddedness distribution in the next section.

%%%%%node - degree figure%%%%%%%%%%%%%%
\begin{figure*}[!t]
%\centerline{\subfloat[Random $k$-tree, $k$=3]{\includegraphics[width
%=1.5in]{rk3_deg}
%\label{fig_rk3_deg}}
%\hfil
%\subfloat[Random $k$-tree, $k$=8]{\includegraphics[width=1.5in]{rk8_deg}
%\label{fig_rk8_deg}}
%\hfil
\centerline{\subfloat[Facebook]{\includegraphics[width=2.1in]{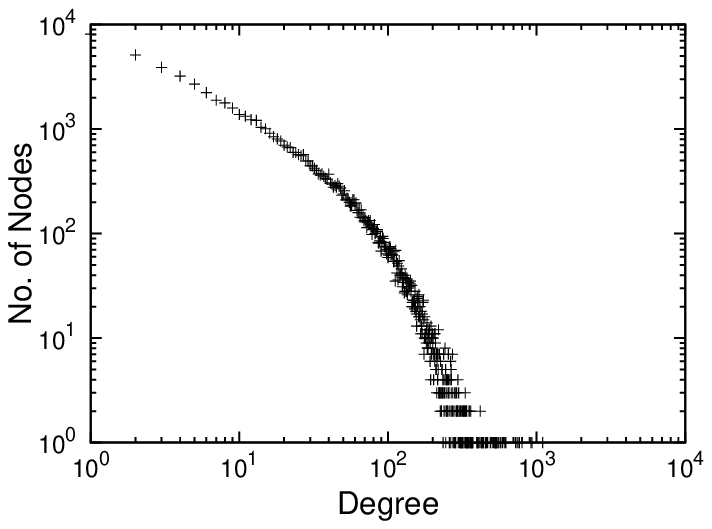}
\label{fig_fb_deg}}
\hfil
\subfloat[Orkut]{\includegraphics[width=2.1in]{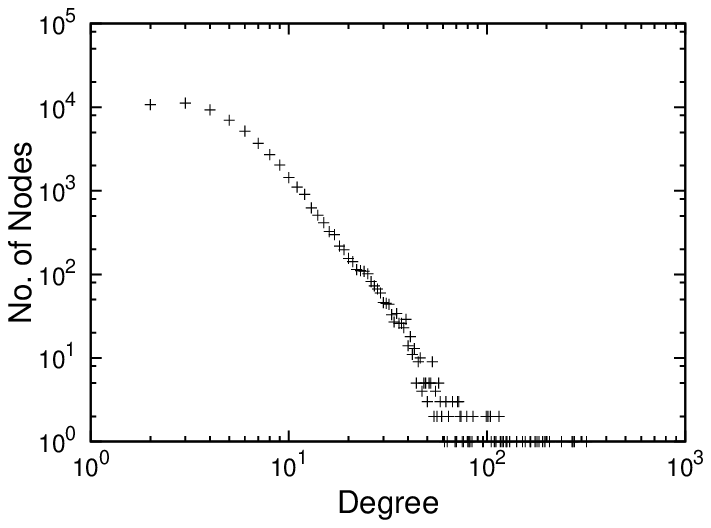}
\label{fig_ork_deg}}}
\caption{Node degree distribution in Facebook and Orkut on a log-log scale. The x-axis represents the vertex degree and the y-axis shows the proportion of nodes having the corresponding
node degree.
}
\label{fig_deg_all}
\end{figure*}

%%%%%%%%% embeddedness figures

\begin{figure*}[thb]
%\centerline{\subfloat[Random $k$-tree, $k$=3]{\includegraphics[width
%=1.5in]{rk3_embed.eps}
%\label{fig_rk3_embed}}
%\hfil
%\subfloat[Random $k$-tree, $k$=8]{\includegraphics[width=1.5in]{rk8_embed.eps}
%\label{fig_rk8_embed}}
%\hfil
\centerline{\subfloat[Facebook]{\includegraphics[width=2.1in]{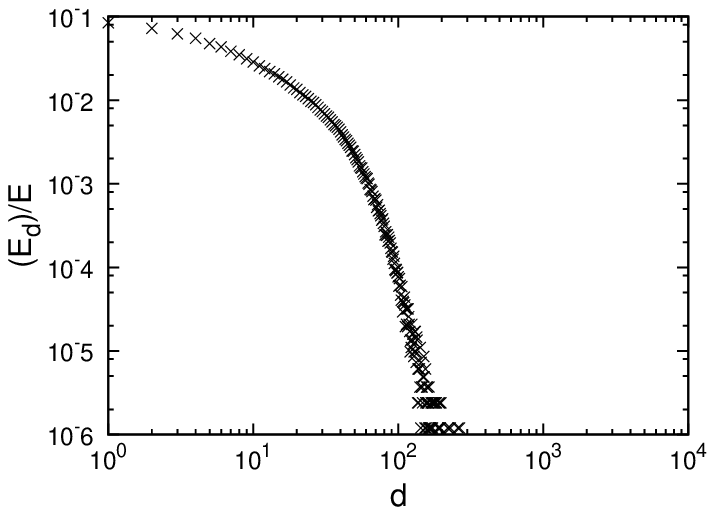}
\label{fig_fb_embed}}
\hfil
\subfloat[Orkut]{\includegraphics[width=2.1in]{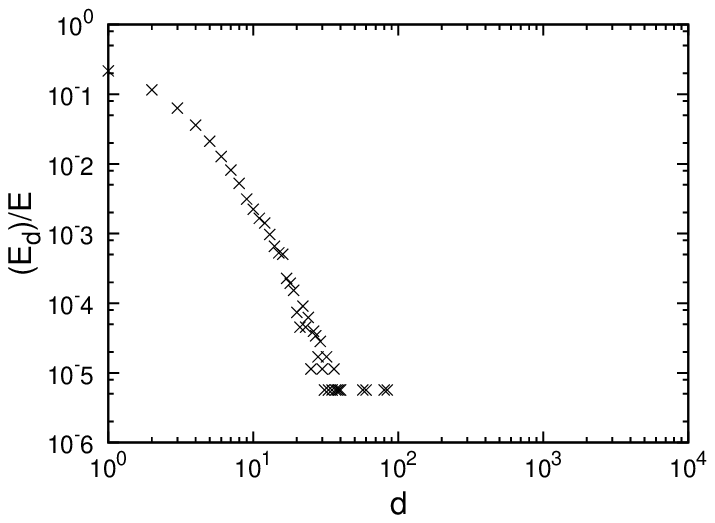}
\label{fig_ork_embed}}}
\caption{Embeddedness distribution in Facebook and Orkut on a log-log scale. The x-axis represents the degree of embeddedness and
the y-axis shows the proportion of edges having the corresponding degree of embeddedness.
}
\label{fig_embed-all}
\end{figure*}

\subsection{Datasets}
\label{sec:data}
We studied the datasets collected from two real-world OSNs: Facebook and Orkut, which are hugely popular OSN services that helps their users to get connected to each other through constant interactions. %Orkut is another popular OSN service provided by Google with a majority of its users base in Asian countries.
Both OSNs have seen tremendous growth in the past few years.
The size of these two OSNs is still growing, making the collection of complete data on these networks extremely
hard, if not impossible. As a result, researchers resort to various methods for collecting a representative sample of these networks.
The datasets for Facebook and Orkut considered in our empirical studies consist of representative samples and were made available to
us by Mislove et al \cite{{mislove-2007-orkut},{viswanath-2009-fb}}.

The Facebook dataset~\cite{viswanath-2009-fb} consists of user links and their wall\footnote{Wall is a space on every user's profile
page that allows friends to post messages to the user.} posts from the New Orleans regional network. The dataset had a total of around
$63$k users with more than $800$k user-to-user links and $870$k wall posts amongst these users for a period of around $3$ years. More details
on the method of extracting the dataset can be obtained from~\cite{viswanath-2009-fb}.

The Orkut dataset~\cite{mislove-2007-orkut} consists of more than $3$ million users.\nop{ with more than $220$ million
user-to-user links.} The sheer size of the dataset as a whole makes it extremely hard and very time consuming to obtain its statistical
results. Hence, we generated a sample of the network for our empirical study. Briefly, the sampling method, based on the
Metropolis-Hasting Random Walk (MHRW) algorithm proposed in~\cite{gjoka-walking}, selects an initial node, $v$, at random and then
proceeds to select the next node, $w$, from the list of its neighbors with a probability of $min(1, \frac{deg(v)}{deg(w)})$, where
$min$ means the minimum value. More subtle details on the sampling method of MHRW could be found at~\cite{gjoka-walking}.
\nop{This process continues iteratively until the stopping criteria is met.}We use this sampling technique since it results in an
unbiased sample of the network when compared to the traditional Breadth First Search (BFS) and Random Walk (RW) techniques~\cite{gjoka-walking}.
After sampling, we obtained a smaller Orkut dataset consisting of around $60$k users with more than $175$k user-to-user links.

\subsection{The Vertex Degree Distribution}
As many real-world networks have been empirically shown to have a power law vertex degree distribution, it is not a surprise
for us to observe roughly a power law vertex degree distribution in the sampled Orkut dataset as shown in Fig.~\ref{fig_ork_deg}.
For the Facbook dataset, we observed that the vertex degree distribution can hardly be called power law  as has been
previously argued~\cite{gjoka-walking}. \nop{\footnote{http://11011110.livejournal.com/179439.html?thread=455919}.}Instead, we can identify two regimes, roughly $[1, 40)$ and $[40, 1098]$, with each approximated by power law exponents
$1.70$ and $2.59$, respectively (Fig.~\ref{fig_fb_deg}). The similar multistage behavior has been observed before in~\cite{gjoka-walking},
although the range of the two regimes and their corresponding exponents are different due to the different
datasets used in the studies in~\cite{gjoka-walking}.

\begin{figure}[h]
\centering
\includegraphics[width=2.1in]{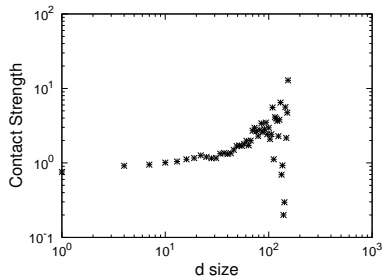}
\caption{Facebook communication pattern and edge embeddedness. The x-axis represents the degree of
edge embeddedness and the y-axis represents the average contact strength of the edges  with the corresponding embeddedness.}
\label{fb_msgs}
\end{figure}

\subsection{The Edge Embeddedness Distribution and Contact Strength of Social Ties}

\begin{figure*}[!t]
\centerline{\subfloat[Random $k$-tree, $k$=3]{\includegraphics[width
=2.1in]{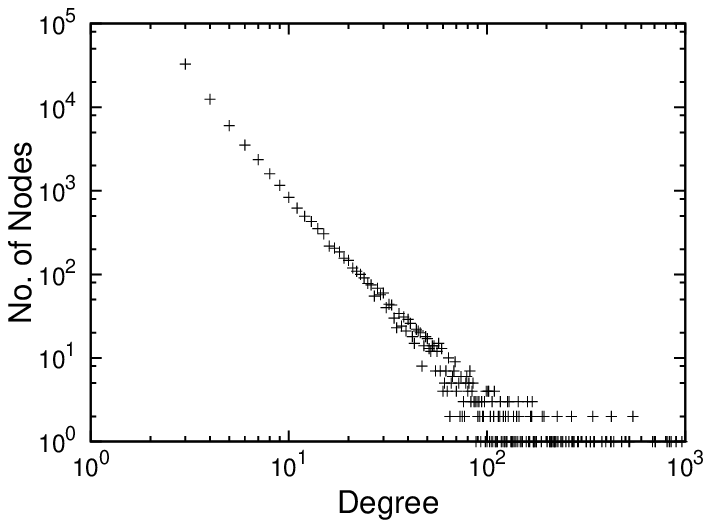}
\label{fig_rk3_deg}}
\hfil
\subfloat[Random $k$-tree, $k$=5]{\includegraphics[width=2.1in]{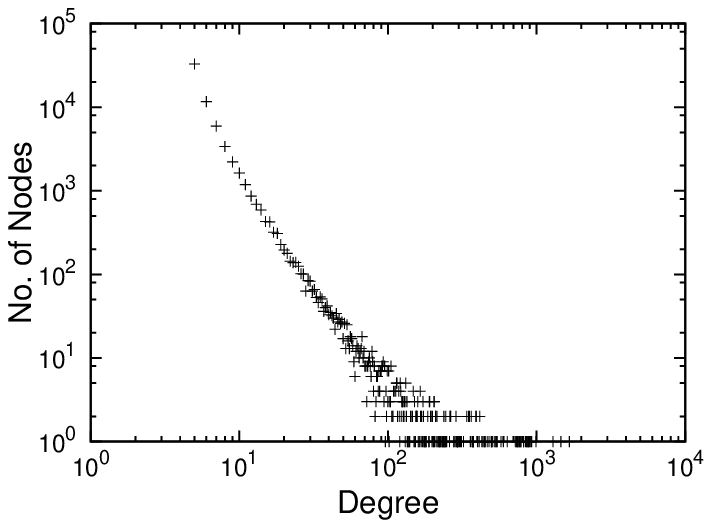}
\label{fig_rk8_deg}}}
\caption{Node degree distribution of random $k$-tree model. The x-axis represents the vertex degree and the y-axis shows the proportion of nodes having the corresponding node degree.}
\label{fig_deg_rk}
\end{figure*}

\begin{figure*}[!t]
\centerline{\subfloat[Random $k$-tree ($k=3$), BA model ($m=3$)]{\includegraphics[width
=2.1in]{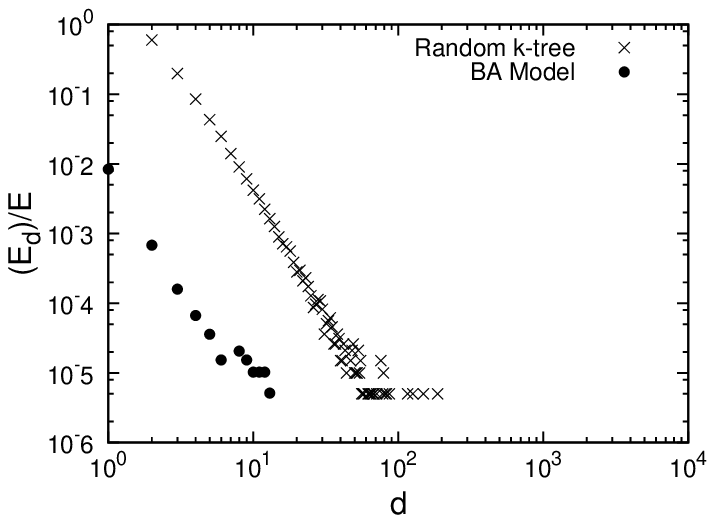}
\label{fig_rk3_embed}}
\hfil
\subfloat[Random $k$-tree ($k=8$), BA model ($m=8$)]{\includegraphics[width=2.1in]{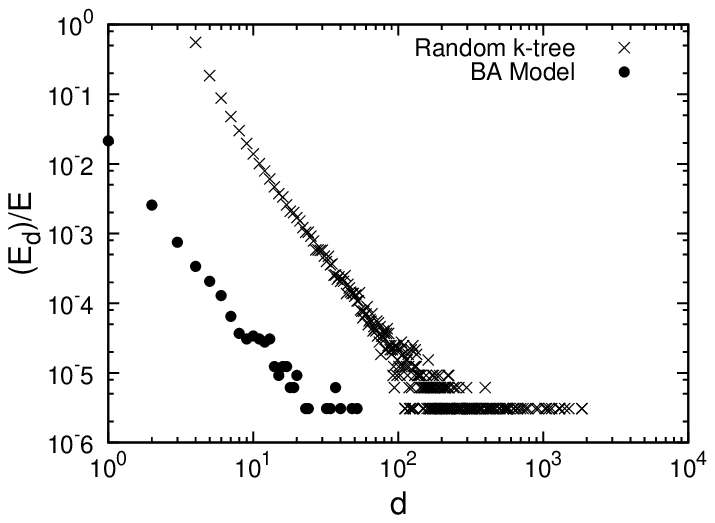}
\label{fig_rk8_embed}}}
\caption{Embeddedness distribution of random $k$-tree model. The x-axis represents the degree of embeddedness and the y-axis shows the proportion of edges having the corresponding degree of embeddedness.}
\label{fig_embed_rk}
\end{figure*}

What intrigued us is that the edge embeddedness distribution in both datasets are found to have a behavior
similar to that of their node degree distribution.
In Figs.~\ref{fig_fb_embed} and~\ref{fig_ork_embed}, we plot the embeddedness distribution in the Facebook and Orkut datasets, respectively.

From these figures it can be observed that the sampled Orkut network tends to have a power law embeddedness distribution
with the power law exponent of $2.91$. The embeddedness distribution in the Facebook dataset does not follow
a power law distribution. Similar to its node degree distribution, we can also roughly identify two regimes
$[1, 50)$ and $[50, 265]$ with  power law exponents around $1.69$ and $3.50$, respectively.
We know of no previous work reporting the distribution of the edge embeddedness of a real-world OSN.

Another interesting observation obtained in our study is the correlation between the embeddedness of the edges  and the contact strength of
the social ties represented by the edges in the Facebook dataset. The \textbf{contact strength} of an edge is defined
as the total number of wall posts posted by the two end nodes of the edge to each other's wall and can be regarded as a metric
measuring the level of communications between the two end nodes.

In Fig. \ref{fb_msgs}, we plot the contact strength of edges as a function of their embeddedness, on a \textit{log-log scale}.
The plot clearly shows that the contact strength increases with the increase of the degree of embeddedness. Even more interesting
to note in Fig.~\ref{fb_msgs} is that a two-stage pattern can also be observed of the contact strength
with the boundary between the two stages coinciding well with the boundary between the two stages in the edge embeddedness distribution
shown in Figure~\ref{fig_fb_embed}.

It is well-known that a heavy-tailed node degree distribution has significant implications on the robustness of an information network.
We believe that the behavior of the embeddedness distributions and its correlation with the contact strength of social ties  as we have observed in the
Facebook dataset are of great significance in many practical applications such as those in~\cite{Zhu09,Ioannidis10}.

\section{Modeling the Embeddedness Distribution  with Random $k$-trees}
\label{Sec:modeling}
While there have been numerous mathematical models in the literature, most notably the well-known BA model,
designed to model the power law node degree distribution
observed in real-world networks, none of them has been proved theoretically or shown empirically to have a power law embeddedness distribution.

In fact, our simulations show that in networks generated with the preferential-attachment-based models, the number of triangles is too low to draw any
meaningful empirical observations on the edge embeddedness distribution. Refer to Fig.~\ref{fig_embed_rk} for a comparison of the edge embeddedness distributions of networks generated from the BA model and the random $k$-tree model with the same edge density.
We also note that none of these previous models can be used to model  the multi-stage degree distributions and embeddedness distributions.

In an effort to search for a good generative model for the power law edge embeddedness distribution, we found out that the random $k$-tree not only has a power law degree distribution as established in \cite{Gao09}, but also has a power law edge embeddedness distribution. A formal proof of the power law embeddedness distribution will be provided in the next section.

Fig.~\ref{fig_deg_rk} shows a \textit{log-log plot} of the node degree distribution in networks
generated from the random $k$-tree model with different values of $k$. We see that the node degree distribution of random $k$-tree model follows
a power law distribution, a fact that has been mathematically established in~\cite{Gao09}.

In Fig.~\ref{fig_embed_rk}, we compare the embeddedness distribution of networks generated from the random $k$-tree model
and the BA model with the same edge density. From the figures, we can infer that the distribution
of edge embeddedness in the random $k$-tree model follows a power law,
which will further be proved in Section~\ref{Sec:proof} of this paper. On the other hand, it is evident that the BA model fails to capture the richness of the edge embeddedness distribution simply because the number of triangles in the networks it generates
is too low.
%The above results demonstrate the strong similarities between the Orkut dataset and graphs generated with the random $k$-tree model.

As has been discussed in the previous section, the degree distribution and the embeddedness distribution of Facebook
have a behavior different from that of the random $k$-tree model.
To understand the two-stage degree and embeddedness distributions observed in the Facebook dataset,
we made the following assumption:
\begin{quotation}
\noindent\textit{There are several types of users, each with a different social connection behavior. As the network evolves over time,
when a user joins the network, he may create social ties  to other users of a different type  as well as social ties
to users of its own type. While the vertex degree and the edge embeddedness of a given type of users viewed in isolation may
have a power law behavior, it is the aggregate effect of users of different types that results in the observed
two-stage (or multi-stage) power law distributions shown in Figs.~\ref{fig_fb_deg} and~\ref{fig_fb_embed} .}
\end{quotation}

\nop{
\begin{center}
\begin{minipage}{8cm}
\noindent
 \textit{There are several types of users, each with a different social connection behavior. As the network evolves over time,
when a user joins the network, he may create social ties  to other users of a different type  as well as social ties
to users of its own type. While the vertex degree and the edge embeddedness of a given type of users viewed in isolation may
have a power law behavior, it is the aggregate effect of users of different types that results in the observed
two-stage (or multi-stage) power law distributions shown in Figures~\ref{fig_fb_deg} and~\ref{fig_fb_embed} .}
\end{minipage}
\end{center}
}
We have to emphasize that the validity of this assumption, just as the assumption made when Barab\'{a}si and Albert \cite{barabasi-1999-286}
proposed their preferential attachment model, needs to be further verified in a variety of OSNs and we leave it as an interesting future work.

Based on the above assumption, we propose the following mixed random $k$-tree model to model the phenomena:
\begin{definition}
\textbf{Mixed random $k$-tree model} is a variant of the random $k$-tree model by mixing different $k$ values
in the $k$-tree process. Formally, given two integers $k_1, k_2 (k_1< k_2)$ and starting with an initial
$k_2$-clique $\rktree{k_2}{n}$, we construct a sequence of graphs $\{\rktree{k_i}{n}, n \geq k_i\}$ by adding vertices
to the graph one at a time, where $k_i$ is a randomly chosen integer in the range of $[k_1, k_2]$ with a pre-defined
probability $p_i (\sum_{i=k_1}^{k_2}p_i =1)$. When a new vertex $v_{n + 1}$ is added, it is connected to the $k_i$
vertices of a $k_i$-clique selected uniformly at random from all the $k_i$-cliques in the previous graph.
\end{definition}
The intuition is that in the mixed random $k$-tree model, all nodes assigned to the same value of $k$ when joining the
network are of the same user type. When viewed in isolation, vertices of the same type evolve in exactly the same way as
they do in a pure random $k$-tree model. By allowing a vertex of a given type to be adjacent to vertices of another type
(and thus contribute to the degree and embeddedness of vertices of that type), an overall multi-stage power law distribution
emerges.

\begin{figure}[hpt]
\centering
\includegraphics[width=2.1in]{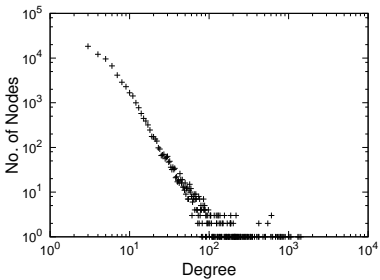}
\caption{Node degree distribution of mixed random $k$-tree, $k= 3$ to $12$.}
\label{fig_deg_mixrk}
\end{figure}

\begin{figure}[hpt]
\centering
\includegraphics[width=2.1in]{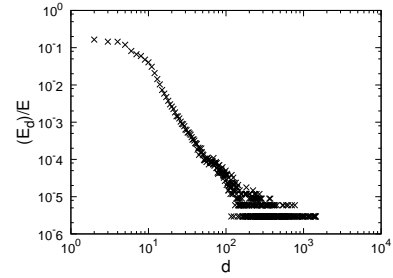}
\caption{Embeddedness distribution of mixed random $k$-tree, $k= 3$ to $12$. The x-axis represents the degree of embeddedness and the y-axis shows the proportion of edges having the corresponding degree of embeddedness.}
\label{fig_emb_mixrk}
\end{figure}

Figs.~\ref{fig_deg_mixrk} and~\ref{fig_emb_mixrk} show the node degree distribution and the embeddedness distribution,
respectively, in the graphs generated with the mixed random $k$-tree model, using parameters $k= 3$ to $12$ (i.e., $k_1=3, k_2=12$) and
preset probabilities of $0.30, 0.20, 0.16, 0.11, 0.06, 0.05, 0.04, 0.03, 0.03, 0.02$, respectively. From the figures, the mixed random $k$-tree model can generate graphs having multi-stage statistics similar to that of Facebook. We note that by adjusting the parameters, we could obtain different multi-stage statistics. We leave the question of how to
choose the parameters to fit the behavior of a particular real-world network as an interesting future work.
%The phenomenon is easy to understand because
%the random $k$-tree model with a constant $k$ value generates a power law distribution of node degree and a power law
%distribution of embeddedness in the whole degree range (i.e., single stage). The above parameters are chosen to approximate
%the multistage statistics of Facebook.

We remark that other random models, such as the BA model and its variants, can also be modified to define a mixed random model to capture
the phenomenon of a multi-stage node degree distribution. But our simulations indicate that these mixed variants of the BA model
fail to capture the rich statistical behavior of the edge embeddedness in OSNs simply because of the extremely low number of
triangles in the networks they generate.

\section{Power Law Distribution of Embeddedness of Random $k$-Trees}
\label{Sec:proof}
%Comments: The proof of power law distribution of embeddedness in random k-tree.
In this section, we prove the following theorem to establish the power law distribution of
the edge embeddedness of a random $k$-tree.
\begin{theorem}
\label{thm-embed-law}
Assume that $k > 2$.
In the random $k$-tree process $\{\rktree{k}{n}, n \geq k\}$, the proportion of the edges $e$
with edge embeddedness $\degree{\rktree{k}{n}}{e} = d$ has the following power law distribution with high
probability\footnote{In the theory of random graphs, by ``with high probability"
we mean that the probability of an event tends to 1 as the size of the graph tends to infinity. All the existing
mathematical results on the power law degree distribution of models for complex networks are established in this form.}:
\begin{equation}
\label{eq-embed-law}
d^{-(1 + \frac{k}{k - 2})}.
\end{equation}
\end{theorem}

To begin with, we first consider the number of $k$-cliques containing a particular edge.
Let $e = \{u, v\}$ be an edge and assume w.l.o.g that $u$ is added before $v$, i.e., the edge
$e$ is ``born" when $v$ is added to $\rktree{k}{n}$.  We have the following observation:
\begin{lemma}
\label{lem-edge-neighbor}
Let $k > 2$ and $c_e^*$ be the number of $k$-cliques that contain the edge $e$. Then,
$$
 c_e^{*} = \binom{k - 1}{k - 2} + \binom{k - 2}{k - 3} (\degree{\rktree{k}{n}}{e} - (k - 1)).
$$
\end{lemma}
\begin{proof}
When $e$ is created as a result of adding the vertex $v$, exactly $\binom{k - 1}{k - 2} = k - 1$ $k$-cliques containing $e$ are created.
For each vertex $w$ added after $v$ is in the network, new $k$-cliques containing $e$ are created if and only if
$w$ is made adjacent to both $u$ and $v$ (and consequently, $\degree{\rktree{k}{n}}{e}$
increased by $1$. If this occurs, exactly $\binom{k - 2}{k - 3} = k - 2$ new $k$-cliques are created
that contain the edge $e$.

Note that the edge embeddedness of $e$ is initially $k - 1$ when $e$ is created. This is because when $v$ is added to the graph, it is made adjacent to a $k$-clique that contains $u$. Therefore, the number of newly-added vertices that form a triangle with $e$
is equal to $\degree{\rktree{k}{n}}{e} - (k - 1)$. The lemma follows.
\end{proof}

Let $\mathcal{C}_n$ be the number of $k$-cliques in $\rktree{k}{n}$. Since every time a new vertex is added
to the network, exactly $k$ different new $k$-cliques are created, the number of $k$-cliques in $\rktree{k}{n}$ is $(n - k)k + 1$, i.e., $\mathcal{C}_n = (n - k)k + 1$. It follows from Lemma~\ref{lem-edge-neighbor} that
given $\rktree{k}{n}$, the probability for the new vertex $v_{n + 1}$
to form a triangle with the two endpoints of an edge $e = \{u, v\}$ is
\begin{eqnarray}
\label{eq-newvertex-prob}
&& \probab{ u \textrm{ and } v \textrm{ are adjacent to } v_{n + 1}}{} = \frac{c_e^{*}}{\mathcal{C}_n} \nonumber \\
&\ \ \ & = \frac{(k - 1) + (k - 2)(\degree{\rktree{k}{n}}{e} - (k - 1))}{(n - k)k + 1} \nonumber \\
&\ \ \ &= \frac{ (k - 2)\degree{\rktree{k}{n}}{e} - b_k}{c_k n}
\end{eqnarray}
where
%$$
%\vcprob{d}{k}{n} = \frac{ (k - 2)d - b_k}{c_k n},
%$$
$b_k = (k - 1) (k - 3)$ and $c_k = k - \frac{k^2 - 1}{n}$. Note that in addition to the constants, the above conditional
probability only depends on $\degree{\rktree{k}{n}}{e}$. With these preparations, we are now ready to prove
Theorem~\ref{thm-embed-law}.

\vspace{0.5cm}
\noindent
{\bf Proof of Theorem 1:}
Let $T_d(n)$ be the number of edges $e$ with edge embeddedness $\degree{\rktree{k}{n}}{e} = d$.
We now derive a system of recursive equations for the expectation $\expectation{T_d(n)}{}$ of $T_d(n)$.
We will focus on the case of $d > k - 1$. The case of $d = k - 1$ is similar.

Let $I_{d}(e, n)$ be the indicator function for the event that the embeddedness of an edge $e$ is
$d$, i.e.,
$$
I_d(e, n) = \left \{
\begin{array}{ll}
1, & \ \ \degree{\rktree{k}{n}}{e} = d \\
0, & \ \ \textrm{otherwise.}
\end{array}
\right.
$$
Then $T_d(n)$ is the sum of $I(e, n)$'s over all the edges, i.e., $T_d(n) = \sum\limits_{e \in \rktree{n}{k}} I_{d}(e, n)$.
By the definition and properties of conditional expectation, we have
\begin{eqnarray*}
&& \expectation{T_{d}(n + 1)\ |\ \rktree{k}{n}, n \geq k }{} \\
   &=& \expectation{\sum\limits_{e\in \rktree{k}{n}} I_d(e, n + 1)\ |\ \rktree{k}{n}, n \geq k }{} \\
   &=& \sum\limits_{e\in \rktree{k}{n}}\expectation{I_d(e, n + 1)\ |\ \rktree{k}{n}, n \geq k}{}
\end{eqnarray*}
For a particular edge $e$, we see that $\degree{\rktree{k}{n + 1}}{e} = d$ if and only if either one of the following
two cases occurs:
\begin{enumerate}
\item $\degree{\rktree{k}{n}}{e} = d$ and $v_{n + 1}$ does not form a triangle with $e$;
\item $\degree{\rktree{k}{n}}{e} = d - 1$ and $v_{n + 1}$ forms a triangle with $e$.
\end{enumerate}
So, if we write $\vcprob{d}{k}{n} = \frac{ (k - 2)d - b_k}{c_k n}$,
we have from Equation (\ref{eq-newvertex-prob}) that
\begin{eqnarray*}
 &&\expectation{I_{d}(e, n + 1)\ |\ \rktree{k}{n}, n \geq k}{} \\
 && = \vcprob{d - 1}{k}{n}I_{d - 1}(e, n) + (1 - \vcprob{d}{k}{n})I_{d}(e, n).
\end{eqnarray*}
Therefore,
\begin{eqnarray*}
&&\expectation{T_{d}(n + 1)\ |\ \rktree{k}{n}, n \geq k}{} \nonumber \\
&&= \vcprob{d - 1}{k}{n}T_{d - 1}(n) + (1 - \vcprob{d}{k}{n})T_{d}(n).
\end{eqnarray*}
Taking unconditional expectation on both sides in the above equation, we have from the properties
of conditional expectation that
\begin{equation}
\expectation{T_{d}(n + 1)}{} = \vcprob{d - 1}{k}{n}\expectation{T_{d - 1}(n)}{}
  + (1 - \vcprob{d}{k}{n})\expectation{T_{d}(n)}{}.
\end{equation}
Based on the above recursion, it can be proved by induction that
for any small constant $\epsilon > 0$, there exists a constant $n_{\epsilon}$
such that for all $n > n_{\epsilon}$, we have
$$
\expectation{T_{d}(n)}{} = \beta_{d}n + \epsilon,
$$
namely, $\expectation{T_{d}(n)}{} = \beta_{d}n + O(1)$,
where $\beta_{d}$ satisfies the following simple equation
$$
  \beta_{d} = \frac{a_k(d - 1) - b_k}{a_kd - b_k + c_k}\beta_{d - 1}.
$$
where $a_k = k - 2$, $c_k = k$, and $b_k = (k - 1)(k - 3)$. The unique solution for the simple
recursive equation for $\beta_d$ is
$$
 \beta_d = \prod\limits_{i = k - 1}^{d}\frac{a_k(i - 1) - b_k}{a_ki - b_k + k}
   = \frac{\Gamma(d - \frac{b_k}{a_k})}{\Gamma(d - \frac{b_k}{a_k} + \frac{k}{a_k} + 1)}.
$$
By Stirling's approximation for the Gamma function, $\beta_d$ is asymptotically equivalent to
$
d^{-(1 + \frac{k}{k - 2})}.
$

Since there are $kn$ edges in a random $k$-tree, we see that the average proportion $\frac{1}{kn}\expectation{T_{d}(n)}{}$
of the number of edges with embeddedness $d$ is asymptotically $d^{-(1 + \frac{k}{k - 2})}$.
By applying Azuma's Inequality~\cite{Mitzenmacher}, it can be shown that the proportion $\frac{1}{kn}T_d(n)$ of the edges with embeddedness $d$
is equivalent to
$
d^{-(1 + \frac{k}{k - 2})}
$
with high probability. We omit the details here due to space limit. However, a similar argument for the case of power law node degree distribution could be found in \cite{Gao09}.
This completes the proof of Theorem~\ref{thm-embed-law}. \qed

For random 2-trees, \nop{ it is interesting to note that while the vertex
degree distribution of a random $2$-tree has been shown to have a power law~\cite{Gao09},} we have the following
theorem showing that its edge embeddedness follows an exponential law. Its proof is omitted due to page limit.
\begin{theorem}
The distribution of the edge embeddedness of a random $2$-tree follows  the exponential law $3^{-d}$.
\end{theorem}
\nop{
\begin{proof}
 For the case of $2$-tree, the term $\vcprob{d}{k}{n})$ defined in the proof of Theorem~\ref{thm-embed-law} becomes
$\vcprob{d}{k}{n}) = \frac{1}{2n - 3}$. Therefore the recursive equation for the expected number of
edges with embeddedness $d$ is
$$
\expectation{T_{d}(n + 1)}{} = \frac{1}{2n - 3}\expectation{T_{d - 1}(n)}{}
  + (1 - \frac{1}{ 2n - 3})\expectation{T_{d}(n)}{}.
$$
By induction, it can be established that
$$
\expectation{T_d(n)}{} = 3^{-d}n + O(1).
$$
\end{proof}
}
Finally, we \nop{mention}note that Theorem~\ref{thm-embed-law} can be generalized to the case of
``embeddedness" of small-sized cliques. Let $\degree{G}{C}$ be the number of common neighbors of the vertices in
a clique $C$. We have the following theorem whose proof is omitted to save space. 
\begin{theorem}
\label{thm-embed-clique-law}
For the random $k$-tree $\rktree{k}{n}$, the proportion of $h$-cliques $C$ with $\degree{\rktree{k}{n}}{C} = d$ follows
the power law distribution
$
  d^{-(1 + \frac{k}{k - h})}.
$
where $h < k$ is a constant.
\end{theorem}

\section{Random Partial $k$-Trees and Size distribution of $k$-Clique Communities}
Besides edge embeddedness, communities and their structures have drawn much interest in recent years.
%A large variety of algorithms have been proposed to detect communities in OSN's \cite{Fortunato10}.
There is also a continuing effort to find better definitions for a network community~\cite{Palla05,Falzon00,Mishra08}.
Recently, Palla et al. \cite{Palla05} introduced the notion of a $k$-clique community. As defined in Section II,
a $k$-clique community is a collection of $k$-cliques where every pair of $k$-cliques can be reached from each other
through a sequence of $k$-cliques that share $k - 1$ vertices. One of the intriguing findings in the study of Palla et al. \cite{Palla05}
is that the size distribution of the $k$-clique communities follows a power law in several real-world networks such as the
co-authorship networks, word-association networks, and the protein interaction networks.

%The set of triangles in a $d$-triangle form a special form a 3-clique community and so the measure of k-clique community size can be
%thought of as a generalization to edge embeddedness.
Toivonen et al. \cite{toivonen09social} refer to $k$-clique communities as \emph{clusters}.
They compare a number of random network models for their parameter values of some higher-order structures under the umbrella of
generating random networks that match the number of nodes and edges to real-world network examples.

In this section, we show that a simple variant of the random $k$-tree model is able to capture the characteristic of
the community structure much better than other existing models such as the BA model.

\begin{definition}\textbf(Random Partial $k$-Trees)
A partial $k$-tree is a subgraph of a $k$-tree. A random partial $k$-tree $\rktree{k}{n, r}$ is a graph obtained by
removing uniformly at random $r$ edges in a random $k$-tree $\rktree{k}{n}$.
\end{definition}

\begin{figure}[hpt]
\centering
\includegraphics[width=2.1in]{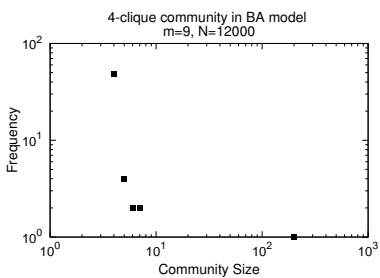}
\caption{$k$-clique communities in graphs created with the BA model}
\label{fig_bamodel}
\end{figure}

\begin{figure}[hpt]
\centering
\includegraphics[width=2.1in]{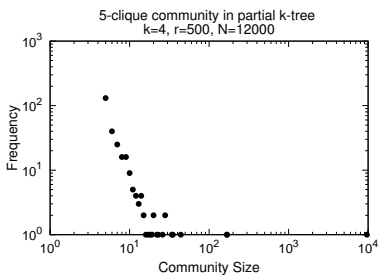}
\caption{Power law community size in partial $k$-trees}
\label{fig_ktree}
\end{figure}

We use the \emph{clique percolation method} of Palla et al. \cite{Palla05} to find the $k$-clique
community sizes in a number of randomly generated networks. We analyze the $(k+1)$-clique community sizes of $\rktree{k}{n, r}$
with various values for $r > 0$. We compare BA model networks and random partial $k$-trees with identical densities: that is,
we set the parameter $m$ in the BA model to the parameter $k$ in partial $k$-trees and keep $r$ relatively small.

Toivonen et al.'s study \cite{toivonen09social} uses two datasets: the \emph{lastfm} network (www.last.fm) of edge density $4.2$  and
an email communication network of edge density $9.6$. Their experiments indicate  that some variants of the BA model are able to
model the size distribution $4$-clique communities in the \emph{lastfm} while all the models they considered have difficulties in
creating a sufficiently large number of $5$-clique communities \cite{toivonen09social}.

Our experimental results confirm  the findings of Toivonen et al. As depicted in
Fig.~\ref{fig_bamodel}, the network generated by the BA model with similar edge density ($m=9$) revealed no power law size distribution of $4$-clique communities. In order to find any nontrivial $5$-clique
community structure, we had to adjust $m$ to $10$ or higher. Nevertheless, even increasing the density to $m=9$
yields no significant $5$-clique community structure.

On the other hand, as shown in Fig~\ref{fig_ktree}, networks generated by the random partial  $4$-tree model reveal a clear power law size
distribution of $5$-clique communities. Further experiments on  our random partial $k$-tree model  $\rktree{k}{n, r}$ with
$k = 9$ and $n = 2000$ shows that the random partial $k$-tree model can generate networks with non-trivial $s$-clique communities with $s$
up to $9$ and  has great potentials to model real-world datasets with higher edge density such as the email dataset used by
Toivonen et al.'s study \cite{toivonen09social}. We omit these experiment results due to the page limit. 

\nop{We will not report the details of the experiments here due to space limit and due
to the relatively small network size $n = 2000$ we used. We note that $n = 2000$ is the largest network size the \emph{clique percolation method} of
Palla et al. \cite{Palla05} can handle for random partial $9$-trees. Due to the structural property of $k$-trees, we believe better data
structures and algorithms can be designed to study the structure of its $s$-clique communities for larger $s$ and will leave it as a future task.}

In conclusion, we observe that the partial $k$-tree model produces community distributions similar to those observed in real-world
networks and such community shapes can be tuned with the parameter $r$. On the other hand, the BA model networks do not reveal
any nontrivial community structure unless the density measure is increased beyond a reasonable threshold.

\section{Related Work}

Research in OSNs has been steadily increasing in the past decade. Analysis of huge on-line social networks~\cite{{1519089},{1242685}} shows that some networks have a scale-free behavior~\cite{barabasi-1999-286} and also exhibit small-world properties~\cite{j1998collective}. There have been numerous mathematical models proposed to model the power law node degree distribution in real-world networks. The well-known one is probably the BA model~\cite{barabasi-1999-286}, which stimulates many similar preferential-attachment-based methods~\cite{Durrett07}. Nevertheless, none of them has been proved theoretically or shown empirically to have a power law embeddedness distribution. 

In the traditional social network literature, the so-called exponential random graph model is also well-known. An exponential random graph model
is specified by a distribution from an exponential family of  distributions over the space of all networks
\cite{jonasson99jap,snijders06sociamethod}. 
An exponential random graph model does have model parameters for higher-order structures and it is possible to use some sophisticated
statistical technique to estimate these parameters. Nevertheless, the exponential random model has its own deficiency. First, the model is
not generative; in fact, generating a random sample from the distribution is a highly non-trivial task. Second, mathematical results
have been established showing that for many parameter settings, as the generated network gets larger, the exponential random graph model
degenerates and becomes trivial in the sense that
it only produces the complete network containing all possible links or networks similar to those generated from the pure Erd\"{o}s-R\'{e}nyi random graphs
(\cite{jonasson99jap,snijders06sociamethod}).

In~\cite{Gao09}, Gao proposed the random $k$-tree model. The work in~\cite{Gao09}, however, only focuses on the power law distribution of node degree distribution of random $k$-trees. The higher-order statistics such as edge embeddedness and community size has not been touched. 

There has been a lot of work in the literature concerning the presence and the significance of edge embeddedness. \nop{For example, embeddedness has been studied in the literature~\cite{gran1985}, where Granovetter explains about the trust between two people when they are connected by mutual friends.}Embeddedness has been utilized in many applications~\cite{Zhu09, Ioannidis10,Wolf09}. It has also been employed to defend against attacks in distributed systems~\cite{1159945} and to detect and prevent email spam~\cite{1267702}. SPROUT~\cite{ilprints763} is a DHT routing algorithm that uses the embeddedness in an OSN to find reliable routes. %In~\cite{Leskovec10signednetworks}, embeddedness is considered in the study of people's relationship and how it affects the structure of OSNs. 
Nevertheless, the above works are intended to better utilize the embeddedness in practice. Our paper lays a solid theoretical foundation for the above work, with which distribution of embeddedness can be modeled and mathematically analyzed.   

\section{Conclusion}
\nop{The Internet has become a social communication and networking platform for people of all ages around the world,
evidenced by the huge success and popularity of OSNs such as Facebook, MySpace, and Twitter. While the original
goal of OSNs is to help people easily interact with their family and friends, and even strangers who share the
same interests or similar profiles, OSNs have evolved from online virtual communities to important service platforms.
This special type of service platforms has seen its adoption in both benign and malicious applications.
It is not uncommon that some universities use OSNs to advertise their educational programs; it is not surprise either
that computer virus designers have taken advantage of OSNs to make virus propagation faster and more effective.
Either way, a better understanding on the statistical behavior of this special type of service platforms is of great importance.

Most existing studies on the statistical behavior of OSNs are based on experimental test over real-world dataset~\cite{Wolf09,Palla05,mislove-2007-orkut,viswanath-2009-fb}.
While empirical studies are valuable, their pitfalls are obvious: the data collection is time consuming; the size of dataset
is usually huge and very hard to handle; the cost of human resources on analyzing and processing the data is nontrivial.
As mathematical models can greatly alleviate the above problems, the call for new mathematical model that is powerful and
flexible enough to approximate the statistical behavior of OSNs, especially the higher-order statistics like the embeddedness,
becomes unprecedentedly urgent.}

In this paper, we showed that real-world OSNs have rich higher-order structural statistics of practical significance that cannot be modeled
by well-known generative models such as the BA model. We studied a newly proposed random $k$-tree model and its different variants,
and showed by both theoretical analysis and empirical simulations that these $k$-tree based models can be used to model not
only the node degree distribution but also higher-order statistics such as the embeddedness and the size of communities of OSNs.
The random $k$-tree model and its variants can be used to easily generate large graphs with statistical features similar to real-world
OSNs such as Facebook and Orkut, and more importantly its unique structure lends itself to amenable mathematical analysis
for higher-order statistics. To the best of our knowledge, this paper is the first that has proved the power law distribution
of embeddedness with random $k$-tree model.

As the norm of mathematical modeling by George Box~\cite{Box87}, ``all models are wrong, but some are useful." While we should not expect that the random $k$-tree model and its variants capture all statistical features of OSNs, we believe, with empirical study and rigorous mathematical proof, that they are extremely useful in the study and utilization of OSNs.


%\section*{Acknowledgment}

    \bibliographystyle{IEEEtran}
    \bibliography{UvicThesis}

\end{document}